\documentclass{aastex701}

\shorttitle{The Spitzer Data Fusion}
\shortauthors{Mattia Vaccari}
\submitjournal{RNAAS}

\begin{document}

\title{The Spitzer Data Fusion - A Far-Ultraviolet to Far-Infrared Multi-Wavelength Database in Spitzer Extragalactic Survey Fields}

\author[orcid=0000-0002-6748-0577,sname='Vaccari']{Mattia Vaccari}
\affiliation{Inter-University Institute for Data Intensive Astronomy (IDIA), Department of Astronomy, University of Cape Town, 7701 Rondebosch, Cape Town, South Africa}
\affiliation{University of the Western Cape, 7535 Bellville, Cape Town, South Africa}
\affiliation{INAF - Istituto di Radioastronomia, via Gobetti 101, 40129 Bologna, Italy}
\email[show]{mattia@mattiavaccari.net}

\begin{abstract}
I present the Spitzer Data Fusion, a multi-wavelength photometric database providing far-ultraviolet to far-infrared flux measurements, together with photometric and spectroscopic redshifts, for 4.4 million IRAC-selected sources over 65\,deg$^2$ in eight of the most widely studied extragalactic survey fields. A companion product, the SERVS Data Fusion, provides 2.8 million sources over 18\,deg$^2$ selected from deeper Spitzer warm-mission imaging. Catalogs are band-merged with wavelength-dependent matching radii, astrometrically registered against 2MASS, and distributed as FITS binary tables, one per field, as \href{https://doi.org/10.5281/zenodo.6120913}{DOI: 10.5281/zenodo.6120913} and as CDS/VizieR \href{https://cdsarc.cds.unistra.fr/viz-bin/cat/II/377}{catalog II/377}. The database is intended as a community resource for photometric redshift calibration, spectral energy distribution fitting, sample selection and multi-wavelength cross-identification, and is a natural bridge between the Spitzer legacy fields and ongoing Euclid, Rubin and SKA precursor surveys.
\end{abstract}

\keywords{
  Astronomical catalogs (205) ---
  Surveys (1671) ---
  Infrared astronomy (786) ---
  Galaxy evolution (594) ---
  Infrared galaxies (790)
}

\section{Introduction}

Selection at mid-infrared wavelengths remains one of the most robust ways to assemble large, approximately stellar-mass-selected samples of galaxies over wide areas and out to high redshift, with a $K$-correction that is favourable and slowly varying over much of the relevant redshift range. The extragalactic fields surveyed in the cryogenic mission of the \textit{Spitzer Space Telescope} \citep{Werner2004} with IRAC \citep{Fazio2004} and MIPS \citep{Rieke2004}, and in particular the SWIRE fields \citep{Lonsdale2003}, have accordingly become reference laboratories for galaxy evolution, and have since accumulated a dense and heterogeneous body of ancillary data from the ultraviolet to the radio.

That richness comes at a cost. Each field is covered by dozens of independent surveys, each with its own source extraction, astrometric solution, photometric convention, flag definitions and file format. Building a coherent, band-merged catalog for even a single field involves substantial bookkeeping that is routinely duplicated across research groups, and that is rarely documented in enough detail to be reproducible. The Spitzer Data Fusion addresses this problem by providing pre-merged, astrometrically registered, versioned and citable multi-wavelength catalogs, one per field, built with a uniform and documented procedure \citep{Vaccari2015}\footnote{\url{https://www.mattiavaccari.net/df}}. It is the photometric counterpart of the Spitzer Spectroscopic Data Fusion \citep{Vaccari2026}, which collects and merges spectroscopic redshift catalogs for a partly overlapping set of fields.

\section{Catalog Construction}

\subsection{Fields Covered}

The Spitzer Data Fusion covers eight extragalactic fields totalling 65\,deg$^2$: the six SWIRE fields \citep[ELAIS-S1, XMM-LSS, CDFS, Lockman Hole, ELAIS-N1 and ELAIS-N2;][]{Lonsdale2003}, the Bo\"otes field of the Spitzer Deep, Wide-Field Survey \citep{Ashby2009}, and the Spitzer Extragalactic First Look Survey \citep[XFLS;][]{Lacy2005}. These fields were chosen for the depth and breadth of their ancillary coverage and for their continuing prominence in Herschel, radio, Euclid and Rubin survey programmes.

\subsection{Band Merging and Astrometric Registration}

Sources are required to be detected at IRAC 3.6 or 4.5\,$\mu$m, which guarantees a homogeneous selection and good astrometric accuracy across the database. Band merging proceeds outwards in wavelength with radii matched to the angular resolution of each band: the 3.6 and 4.5\,$\mu$m catalogs are associated with a 1\,arcsec radius and define the positional reference; the 5.8 and 8.0\,$\mu$m catalogs are then matched to those positions within 1.5\,arcsec; and the MIPS 24, 70 and 160\,$\mu$m catalogs within 3, 6 and 12\,arcsec respectively.

Before merging, every input catalog is registered against 2MASS \citep{Skrutskie2006} as a common astrometric frame: median offsets in right ascension and declination are computed using high-significance detections and applied to the whole catalog, which removes the systematic frame-to-frame offsets that otherwise limit the reliability of cross-identification at fixed radius.

\subsection{Ancillary Data}

Ancillary catalogs are associated to the IRAC positions by nearest-neighbour matching within a homogeneous 1\,arcsec radius. They include GALEX ultraviolet photometry \citep{Martin2005}, optical photometry from SDSS and from field-specific ground-based programmes, near-infrared photometry from 2MASS, UKIDSS \citep{Lawrence2007} and VIDEO \citep{Jarvis2013}, and far-infrared photometry from Herschel PACS and SPIRE surveys. Photometric and spectroscopic redshifts are carried alongside the photometry, the latter drawn from the compilation described in \citet{Vaccari2026}.

\section{Data Model and Access}

Catalogs are distributed as FITS binary tables, one per field, within a single ZIP archive (2.0\,GB for the DR1 release of 28 July 2023). Column names encode the originating survey and band, so that the provenance of every measurement is explicit, and the accompanying README file (\texttt{AAAREADME.MAIN}) documents the contributing catalogs, the matching radii, the astrometric corrections applied and the field-specific caveats.

\begin{figure*}[ht!]
\plotone{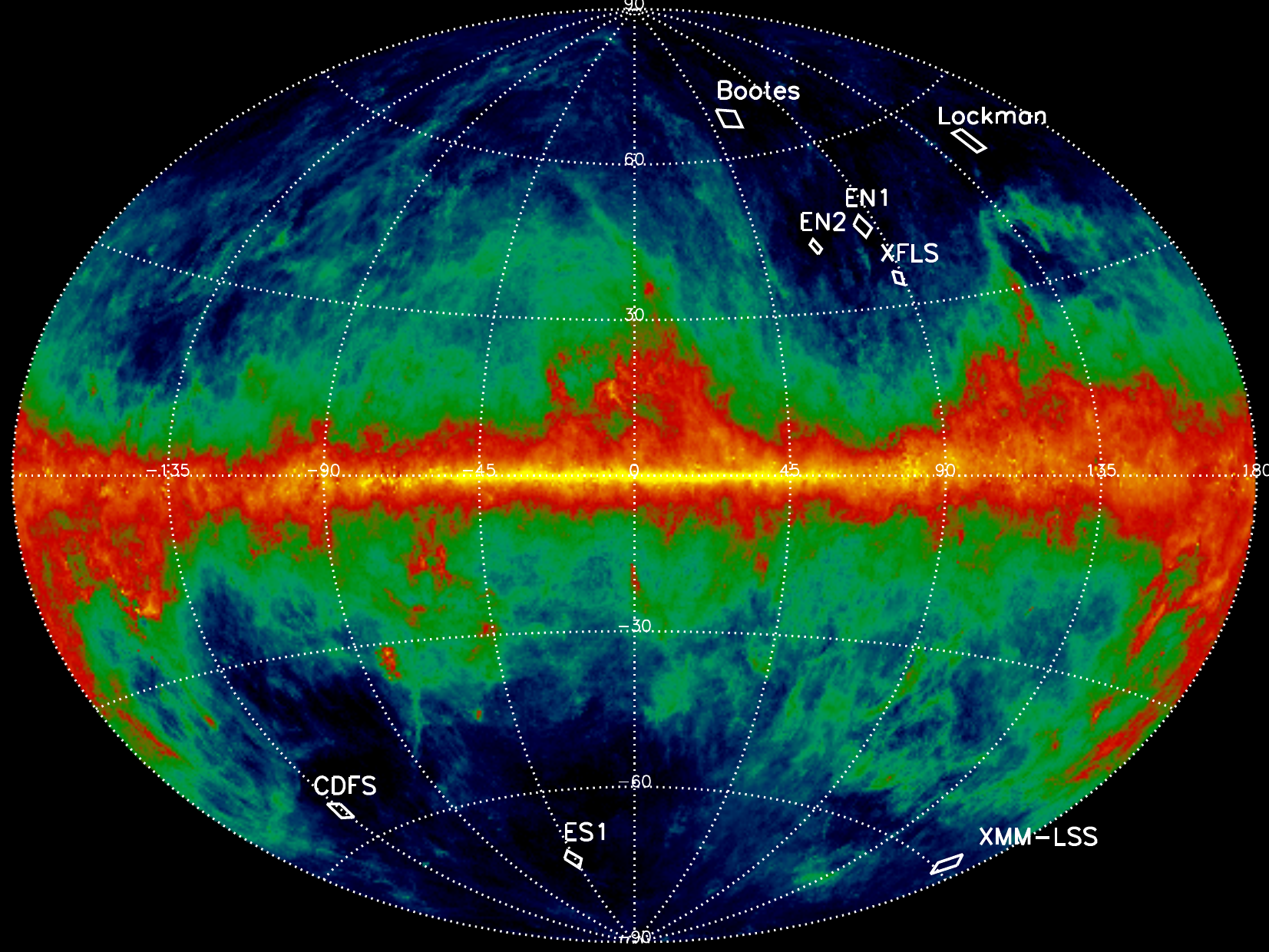}
\plotone{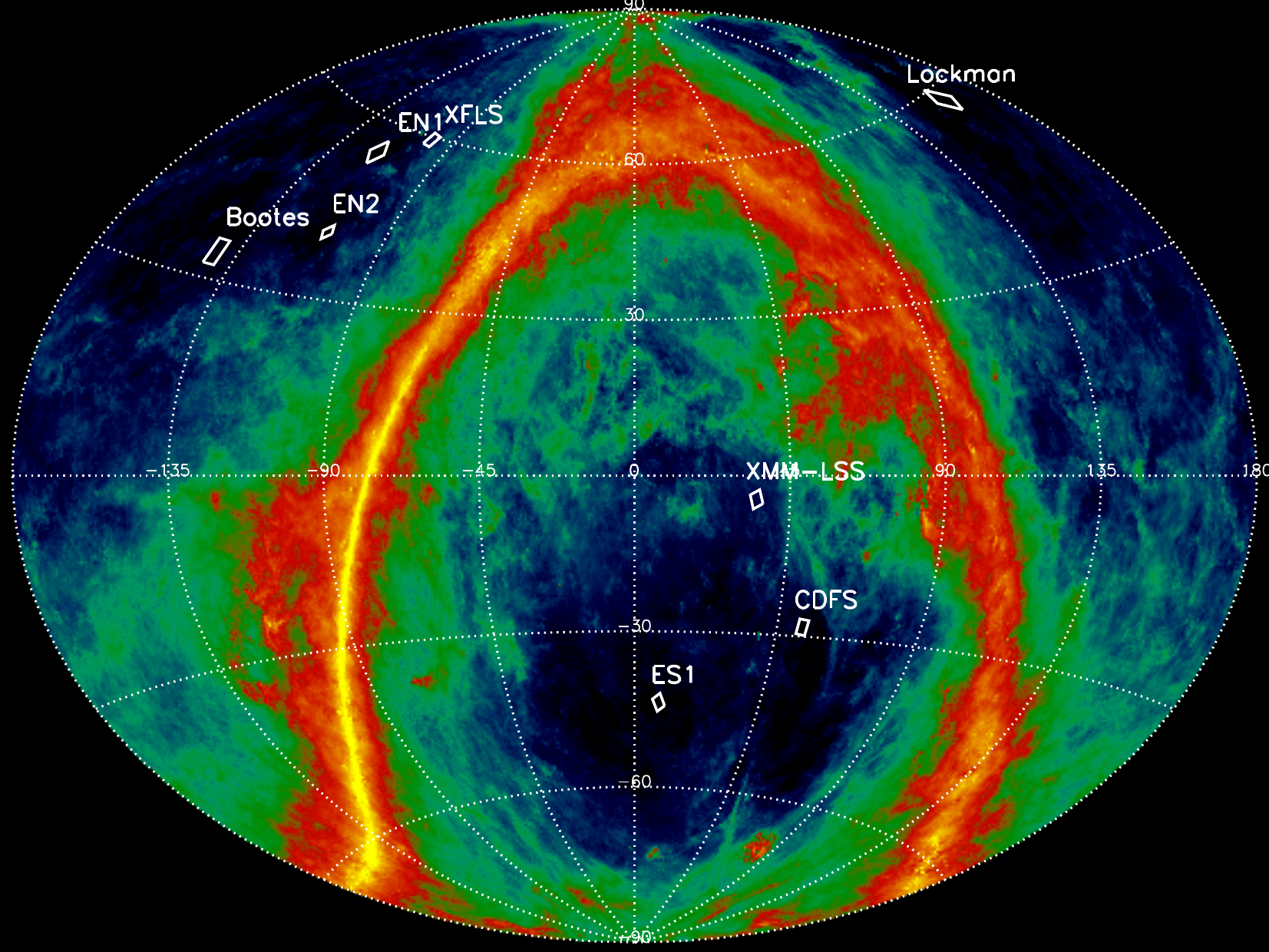}
\caption{Spitzer Data Fusion Sky Coverage in Galactic (Top Panel) and Equatorial (Bottom Panel) Coordinates.}
\label{fig:fields}
\end{figure*}

The database is archived on Zenodo under the concept DOI \href{https://doi.org/10.5281/zenodo.6120913}{DOI: 10.5281/zenodo.6120913}, which resolves to the most recent version, with DR1 archived at \dataset[DOI: 10.5281/zenodo.8192777]{https://doi.org/10.5281/zenodo.8192777} and also served by CDS/VizieR as \href{https://cdsarc.cds.unistra.fr/viz-bin/cat/II/377}{catalog II/377} \citep{2023yCat.2377....0V}. Figure~\ref{fig:fields} shows the sky coverage of the eight fields.

While the Zenodo record includes a selection of data products, the full database is available at \url{https://www.mattiavaccari.net/df/pw/spitzer}.
The MIPS maps and catalogs produced as part of this work are also separately available at \url{https://www.mattiavaccari.net/df/m24} (24 \,$\mu$m) and \url{https://www.mattiavaccari.net/df/mipsge} (70 and 160\,$\mu$m). The companion photometric filter database used for calibration is available at \dataset[DOI: 10.5281/zenodo.7864237]{https://doi.org/10.5281/zenodo.7864237}. Further ancillary data products are also available at \url{https://www.mattiavaccari.net/df}.

\section{The SERVS Data Fusion}

A companion database, the SERVS Data Fusion, applies the same procedure to the deeper 3.6 and 4.5\,$\mu$m imaging obtained during the Spitzer warm mission by SERVS \citep{Mauduit2012}, and delivers 2.8 million IRAC-selected sources over 18\,deg$^2$.
SERVS IRAC12 catalogs are available as \dataset[DOI: 10.5281/zenodo.7929151]{https://doi.org/10.5281/zenodo.7929151}
while the SERVS Data Fusion is available at \url{https://www.mattiavaccari.net/df/pw/servs}.

\section{Science Applications}

The Spitzer Data Fusion has underpinned a range of multi-wavelength galaxy evolution studies. It provided the parent samples and multi-wavelength photometry for infrared luminosity function analyses \citep[e.g.,][]{Vaccari2010,Marchetti2016}, and training and validation sets for photometric redshift estimation of large infrared-selected samples \citep[e.g.,][]{RowanRobinson2013,Pforr2019}. Within the Herschel Extragalactic Legacy Project \citep[HELP;][]{Vaccari2016,Shirley2021} it contributed prior positions and ancillary photometry for the deblending of Herschel maps and for the calibration of photometric redshifts across multiple fields.

Looking ahead, three of the eight fields are Euclid Deep or Auxiliary fields or Rubin Deep Drilling fields, and all are targeted by ongoing or planned radio continuum surveys with LOFAR, GMRT and MeerKAT and by spectroscopic campaigns with 4MOST, MOONS and PFS. The Spitzer Data Fusion is therefore well placed to supply the mid-infrared anchor, the prior source lists and the photometric redshift training sets required to exploit these surveys.

\section{Summary}

I present the Spitzer Data Fusion, a publicly archived multi-wavelength database delivering far-ultraviolet to far-infrared photometry and redshifts for 4.4 million IRAC-selected sources over 65\,deg$^2$ in eight extragalactic survey fields, complemented by the SERVS Data Fusion, which delivers 2.8 million sources over 18\,deg$^2$ from deeper Spitzer warm-mission imaging. Catalogs are band-merged with wavelength-dependent matching radii, registered against a common astrometric frame, fully documented and versioned. The database is available on Zenodo and on CDS/Vizier.

\facilities{Herschel(PACS, SPIRE), Spitzer(IRAC, MIPS), GALEX, Sloan, VST(OmegaCAM), UKIRT(WFCAM), VISTA(VIRCAM), CTIO:2MASS, FLWO:2MASS, IRSA, NED, CDS, Zenodo}

\software{STILTS/TOPCAT \citep{Taylor2005,Taylor2006}}

\bibliographystyle{aasjournalv7}

\begin{thebibliography}{}

\bibitem[Ashby et al.(2009)]{Ashby2009}
Ashby, M.~L.~N., Stern, D., Brodwin, M., et al.\ 2009,
\apj, 701, 428. \doi{10.1088/0004-637X/701/1/428}

\bibitem[Fazio et al.(2004)]{Fazio2004}
Fazio, G.~G., Hora, J.~L., Allen, L.~E., et al.\ 2004,
\apjs, 154, 10. \doi{10.1086/422843}

\bibitem[Jarvis et al.(2013)]{Jarvis2013}
Jarvis, M.~J., Bonfield, D.~G., Bruce, V.~A., et al.\ 2013,
\mnras, 428, 1281. \doi{10.1093/mnras/sts118}

\bibitem[Lacy et al.(2005)]{Lacy2005}
Lacy, M., Wilson, G., Masci, F., et al.\ 2005,
\apjs, 161, 41. \doi{10.1086/432894}

\bibitem[Lawrence et al.(2007)]{Lawrence2007}
Lawrence, A., Warren, S.~J., Almaini, O., et al.\ 2007,
\mnras, 379, 1599. \doi{10.1111/j.1365-2966.2007.12040.x}

\bibitem[Lonsdale et al.(2003)]{Lonsdale2003}
Lonsdale, C.~J., Smith, H.~E., Rowan-Robinson, M., et al.\ 2003,
\pasp, 115, 897. \doi{10.1086/376850}

\bibitem[Marchetti et al.(2016)]{Marchetti2016}
Marchetti, L., Vaccari, M., Franceschini, A., et al.\ 2016,
\mnras, 456, 1999. \doi{10.1093/mnras/stv2717}

\bibitem[Martin et al.(2005)]{Martin2005}
Martin, D.~C., Fanson, J., Schiminovich, D., et al.\ 2005,
\apjl, 619, L1. \doi{10.1086/426387}

\bibitem[Mauduit et al.(2012)]{Mauduit2012}
Mauduit, J.-C., Lacy, M., Farrah, D., et al.\ 2012,
\pasp, 124, 714. \doi{10.1086/666945}

\bibitem[Pforr et al.(2019)]{Pforr2019}
Pforr, J., Vaccari, M., Lacy, M., et al.\ 2019,
\mnras, 483, 3168. \doi{10.1093/mnras/sty3075}

\bibitem[Rieke et al.(2004)]{Rieke2004}
Rieke, G.~H., Young, E.~T., Engelbracht, C.~W., et al.\ 2004,
\apjs, 154, 25. \doi{10.1086/422717}

\bibitem[Rowan-Robinson et al.(2013)]{RowanRobinson2013}
Rowan-Robinson, M., Gonzalez-Solares, E., Vaccari, M., et al.\ 2013,
\mnras, 428, 1958. \doi{10.1093/mnras/sts163}

\bibitem[Shirley et al.(2021)]{Shirley2021}
Shirley, R., Duncan, K., Campos Varillas, M.~C., et al.\ 2021,
\mnras, 507, 129. \doi{10.1093/mnras/stab1526}

\bibitem[Skrutskie et al.(2006)]{Skrutskie2006}
Skrutskie, M.~F., Cutri, R.~M., Stiening, R., et al.\ 2006,
\aj, 131, 1163. \doi{10.1086/498708}

\bibitem[Taylor(2005)]{Taylor2005}
Taylor, M.~B.\ 2005, Astronomical Data Analysis Software and Systems XIV, 347, 29.

\bibitem[Taylor(2006)]{Taylor2006}
Taylor, M.~B.\ 2006, Astronomical Data Analysis Software and Systems XV, 351, 666.

\bibitem[Vaccari et al.(2010)]{Vaccari2010}
Vaccari, M., Marchetti, L., Franceschini, A., et al.\ 2010,
\aap, 518, L20. \doi{10.1051/0004-6361/201014694}

\bibitem[Vaccari(2015)]{Vaccari2015}
Vaccari, M.\ 2015, PoS (EXTRA-RADSUR2015), 027.
\doi{10.22323/1.267.0027}

\bibitem[Vaccari(2016)]{Vaccari2016}
Vaccari, M.\ 2016, ASSP, 42, 71. 
\doi{10.1007/978-3-319-19330-4\_10}

\bibitem[{{Vaccari}(2023){Vaccari}}]{2023yCat.2377....0V}
{Vaccari}, M. 2023, {VizieR Online Data Catalog: Spitzer Data Fusion main
  catalogue (Vaccari, 2023)},, VizieR On-line Data Catalog: II/377. Originally
  published in: 2015fers.confE..27V

\bibitem[Vaccari(2026)]{Vaccari2026}
Vaccari, M.\ 2026, Research Notes of the American Astronomical Society, 10, 118.
\doi{10.3847/2515-5172/ae6a9a}

\bibitem[Werner et al.(2004)]{Werner2004}
Werner, M.~W., Roellig, T.~L., Low, F.~J., et al.\ 2004,
\apjs, 154, 1. \doi{10.1086/422992}

\end{thebibliography}

\end{document}